\documentclass[12pt,leqno]{article}
\usepackage{graphicx}
\usepackage{indentfirst,csquotes}

\usepackage{color}
\usepackage{cite}
\usepackage{amsmath,amssymb,amsfonts}
\usepackage{algorithmic,mathtools,graphicx}
\usepackage{textcomp,lipsum,fancyhdr}
\usepackage{cuted}
\usepackage{subfig}

\title{A new bandwidth-voltage trade off paradigm in low-loss LNOI electro-optic modulators, using equalizer configuration}
\author{Fatemeh Ghavami, Sara Darbari, Mohammad Kazem Moravvej-Farshi
\\
 \small Faculty of Electrical and Computer Engineering, Tarbiat Modares University, Tehran, Iran.}
\begin{document}

{}
\maketitle

\begin{abstract}
As a crucial element for various applications, including optical communications and analog photonic links, electro-optic modulators have a small size, low voltage, and wide bandwidth. 
 Electro-optic modulators based on the thin-film lithium niobate (TFLN) are highly efficient and have a  large modulation bandwidth. 
However, there is a tradeoff between driving voltage and modulation bandwidth.
In this study, we present the electro-optical modulator based on the TFLN platform by using the electro-optic equalizer. Using the electro-optic equalizer causes that the proposed modulator has a broad bandwidth while the driving voltage is not compromised.
In this proposed modulator, to compensate for the velocity mismatch and minimize interference of the electrode with the waveguide, a two-layer metallization method is used.
This proposed modulator results in minimal optical and microwave losses and allows for a large electro-optic bandwidth of up to 300GHz for a 5 mm-long device. 
Additionally, an efficiency of 4.5 V.cm.
\end{abstract}

\section{Introduction}

With the development of telecommunication systems, the need for small footprint,  low loss higher- bandwidth, and high-speed modulation also grows. Meanwhile, the optical modulator is an integral part of optical communication networks \cite{BW2021,AA216,WG2018}. To achieve all the mentioned goals simultaneously, an ideal optical modulator should use the appropriate PIC material\cite{SR2006,JB2006}. Many platforms can be used in PICs. Famous examples include silicon\cite{XQ2005,SC2015}, indium phosphide\cite{OY2017,AM1993}, polymers\cite{KS2015,LM2002}, and plasmonics\cite{HC2018}. Great features such as high scalability and the ability to integrate with CMOS are the advantages of using silicon. Also, low drive voltage (indium phosphide), and high bandwidth (polymer) can be considered as benefits of these platforms. Despite the great progress of these platforms, they cannot be considered as the ideal ones for photonics.  Because the  ideal platform, which has all the performance merits, is impossible  without an excellent electro-optical effect.
 
Lithium niobite is an amazing and excellent material for realizing ultra-fast modulation with low loss and low half-wave voltage\cite{HM2019,WX2019,AA2020,JM2021}. This material is widely used in fast modulators due to its strong electro-optic effect. Traditional modulators are mainly based on bulk lithium niobate waveguides\cite{AA2020,SR2020}. In these waveguides, which are made with the titanium-indiffusion or proton-exchange method, the refractive index contrast between the core and the clad is very low\cite{CS2021,LX2020}. Therefore the size of the optical mode increased, and this causes the half-wave voltage to increase. By improving fabrication techniques, LNOI thin film wafers became commercially available\cite{LH2017,WJ2015}.
The optical mode size of modulators fabricated from LNOI wafers can be  20 times smaller than ones from bulk lithium niobate.
 In Mach Zehender modulators by reducing the size of the optical mode, the gap between the electrodes can be reduced. In this result,  the footprint and voltage are extremely reduced and the speed is increased\cite{ZM2021,WP2020,SS2021}.

The theoretical literature previously mentioned provides a foundation for the electro-optic modulator.  In this modulator, due to the trade-off between "Bandwidth" and "Half wave voltage", we cannot bring the electrodes too close to each other\cite{RA2016,CL2014}. In \cite{YY2022}, by using an equalizer and changing the polarity, a structure has been proposed which is possible to increase the bandwidth while keeping V-pi constant. However, the main challenge, which has not been considered, is the matching of optical signal and microwave signal speed. Actually, it is essential for increasing the bandwidth.

In this paper, we propose a high-speed electro-optic modulator based on LNOI by using the EO equalizer. Actually, for the first time in this structure, we use a designed dual-electrode approach. In this proposed modulator, a waveguide crossing, which is utilized for polarization rotation, is precisely designed and simulated.  In the designed waveguide crossing section, the velocity matching, which increases the modulator speed, is considered. Additionally, to prevent interference between the waveguide crossing section and electrodes, a dual-metal technology is utilized. At last,  a proper crossing has been presented.
This proposed device has a half wave voltage of 4.5V and can also support a bandwidth over 300 GHz.

\begin{figure}
\centering
\includegraphics[width=0.95\linewidth]{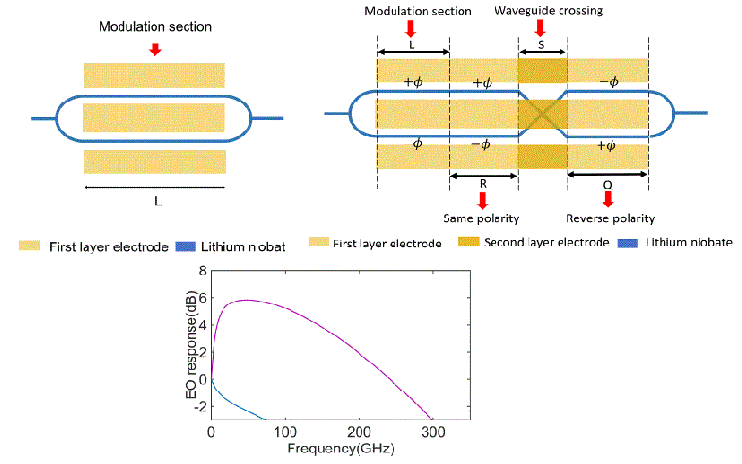}
  \label{fig1_A}
  
\caption{(a) Structure of regular modulator. (b) Structure of proposed modulator.(c) Electro-optic response of modulator with and whithout equalizer}
\label{fig:1}

\end{figure}

\section{Design and simulation}

 \subsection{Theoritical method}

 \subsubsection{Regular modulator}
 
TWM modulators are high-speed modulators that have a good bandwidth. However, due to the trade-off between bandwidth and efficiency, these modulators have speed limitations.
The regular modulator is illustrated in Figure 1a.
This regular modulator is designed and simulated by using COMSOL multiphysics. Additionally, it has very low microwave losses and excellent velocity and impedance matching. but however, this modulator has a bandwidth of 70GHz like Figure 1c.

 As previously noted, there exists a trade-off between  $V_\pi $ and  3dB bandwidth.
However, by using an EO equalizer in the modulator path, we can increase the 3dB bandwidth of the modulator several times without changing the $V_\pi $.
 \subsubsection{Proposed modulator}
The proposed modulator, as illustrated in Figure 1b, is composed of four distinct parts.
The L  section is the main part of the fundamental parts of the modulator where the actual modulation process takes place and shows in Figure 1b. Part R is responsible for incrementing the phase, whereas part S changes the polarity. Finally, in part Q, the negative phase cancels out the phase created by part R.

A modulator gain by using an equalizer can be computed as follow:

Here, ${\phi }_{MZ}$ and ${\phi }_{MZ-with-Equalizer}$ refer to the amount of phase shift that occurs in  regular MZ modulator and  MZ modulator with the equalizer. ${V}_{0}$ is the input voltage and $\alpha_m$ is the microwave loss rate of the modulator. $L$ denotes the length of the modulator without modulator. $R$, $S$, and $Q$ represent the ranges of positive phase, polarization change, and negative polarization, respectively.

Figure3a illustrates the admissible values for $R$ and $Q$. Actually, choosing these parameters can lead to a zero DC gain. by assuming $S=x\mu m$ in the proposed modulator the gain can be calculated.
If the achieved gain is added to the electro-optic response of the regular modulator,  the electro-optic response of the proposed modulator can be shown in Figure 1c.
 As indicated, this proposed modulator can have a 3dB bandwidth exceeding 200 GHz.

\subsection{Device layout}
\begin{figure}
\centering
  \includegraphics[width=1\linewidth]{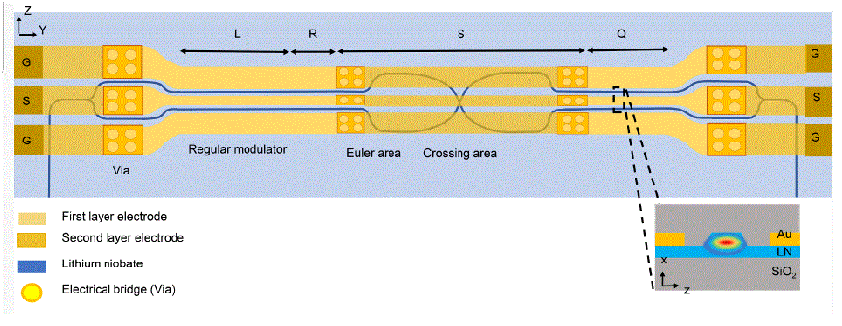}
  \caption{ Structure of proposed modulator. Inset: Cross-sectional modulation area and  Z-Y view of modulator with tow layer metalisation.}
  \label{fig:2}
\end{figure}

Figure \ref{fig:2} illustrates a top view and cross-section of the proposed electro-optic modulator. From bottom to top, the materials used in the device are as follows:
the bottom layer is a substrate made of silicon that has a thickness of x micrometers. On top of the substrate is a x-micrometer-thick buried oxide layer made of silicon dioxide. A thin film layer of Lithium niobate, with a thickness of x nanometers, sits on top of the buried oxide layer. Finally, anotherxmicrometer-thick layer of silicon dioxide acts as the clad layer on top of the Lithium niobate layer.

The optical mode propagates in the Y-direction and due to the fact that the LN wafer is of X-cut, with the highest electro-optic coefficient of Lithium niobate in the Z-direction, electrodes are placed along the Z-axis. If the electrodes were placed along the X-axis instead, the optical mode would not be efficiently modulated by the electric field. Because the optical mode still propagates in the Y-direction.
The electrodes utilized in this study are made of gold and have a thickness of xnm.

According to  Figure \ref{fig:2}, the light in the L region is split into two waveguides. The waveguides travel towards the R region, where the electrodes remain in the first layer. After passing through the R region, the electrodes enter the S region (the crossing region) and are connected to the second layer through a via that is located in the center of the clad.
The thickness of the oxide layer under the electrode in the second layer is greater than  \( x\ \mu\)m to prevent any optical losses in the crossing region. After the crossing region, in the Q region, the electrodes are again placed in the first layer. Finally, the waveguides are combined together. the modulator consists of a waveguide crossing that is formed approximately in the middle of the modulator, with electrodes placed in a GSG(Groung-Signal-Ground) configuration before and after this region in the first layer. On the other hand, in the crossing section, Electrodes are placed in the second layer to avoid interference with the waveguides.

 \subsection{Waveguide crossing design without interference}

In this section, the waveguide crossing design is explained.

As illustrated in Figure 3a., each waveguide in the crossing area comprises three distinct sections: an Euler bend at the beginning of the path, a cross, and an Euler bend at the end of the path.

On the other hand, the microstrip line also consists of three essential parts. The first part is the via, which connects the first metal layer to the second layer. This part has very low losses when considering hollow cylinders. 
The second part is the coplanar waveguide (CPW) is designed. The signal electrode width and the gap between the two electrodes are considered based on a 50 $\Omega$ impedance.

\subsubsection{Via design}
To avoid interference between the waveguide crossing and electrodes, we employ a two-layer metallization process. As a result, as shown in Figure 2, hollow cylindrical vias are utilized to connect the first layer to the second layer. The number and radius of each via are optimally chosen.
By using the HFSS (High-Frequency Structural Simulator) software, $S_{21}$ and $S_{11}$ diagrams are simulated and shown in Figures 1b.

\subsubsection{Waveguide crossing design}
To validate our design and ensure its accuracy, we performed a 3D simulation of the modulator using Lumerical's FDTD software. 
As shown in Figure 3c

\section{Conclusions}

In conclusion, an electro-optic modulator based on lithium niobate along with an equalizer has been proposed. 
By using the electro-ocptic equalizer in this proposed modulator, we can increase the bandwidth without any negative impact on the half-wave voltage.
In the waveguide crossing area of this mentioned equalizer, the dual-layered electrode design has been used to reduce electrode interference with the waveguide. 
Additionally, this mentioned area has been precisely designed and simulated in order to match both optical and microwave signals. 
This is a significant finding as it indicates that this proposed modulator can operate at higher frequencies and with greater efficiency than previously reported systems. 
Also, this proposed modulator has optical losses below 0.1 dB and a bandwidth of over 300 GHz and efficiency of 4.5 V.cm.

\bibliographystyle{unsrtnat}

\begin{thebibliography}{100}

\bibitem{BW2021}
W.~Beardell, B.~Mazur, C.~Ryan, G.~Schneider, J.~Murakowski, and D.~Prather,
  ``Rf-photonic spatial-spectral channelizing receiver,'' \emph{Journal of
  Lightwave Technology}, vol.~40, no.~2, pp. 432--441, 2021.

\bibitem{AA216}
A.~Alvarado, D.~J. Ives, S.~J. Savory, and P.~Bayvel, ``On the impact of
  optimal modulation and fec overhead on future optical networks,''
  \emph{Journal of Lightwave Technology}, vol.~34, no.~9, pp. 2339--2352, 2016.

\bibitem{WG2018}
C.~Wang, M.~Zhang, X.~Chen, M.~Bertrand, A.~Shams-Ansari, S.~Chandrasekhar,
  P.~Winzer, and M.~Lon{\v{c}}ar, ``Integrated lithium niobate electro-optic
  modulators operating at cmos-compatible voltages,'' \emph{Nature}, vol. 562,
  no. 7725, pp. 101--104, 2018.

\bibitem{SR2006}
R.~Soref, ``The past, present, and future of silicon photonics,'' \emph{IEEE
  Journal of selected topics in quantum electronics}, vol.~12, no.~6, pp.
  1678--1687, 2006.

\bibitem{JB2006}
B.~Jalali and S.~Fathpour, ``Silicon photonics,'' \emph{Journal of lightwave
  technology}, vol.~24, no.~12, pp. 4600--4615, 2006.

\bibitem{XQ2005}
Q.~Xu, B.~Schmidt, S.~Pradhan, and M.~Lipson, ``Micrometre-scale silicon
  electro-optic modulator,'' \emph{nature}, vol. 435, no. 7040, pp. 325--327,
  2005.

\bibitem{SC2015}
C.~Sun, M.~T. Wade, Y.~Lee, J.~S. Orcutt, L.~Alloatti, M.~S. Georgas, A.~S.
  Waterman, J.~M. Shainline, R.~R. Avizienis, S.~Lin \emph{et~al.},
  ``Single-chip microprocessor that communicates directly using light,''
  \emph{Nature}, vol. 528, no. 7583, pp. 534--538, 2015.

\bibitem{OY2017}
Y.~Ogiso, J.~Ozaki, Y.~Ueda, N.~Kashio, N.~Kikuchi, E.~Yamada, H.~Tanobe,
  S.~Kanazawa, H.~Yamazaki, Y.~Ohiso \emph{et~al.}, ``Over 67 ghz bandwidth and
  1.5 v v$\pi$ inp-based optical iq modulator with nipn heterostructure,''
  \emph{Journal of lightwave technology}, vol.~35, no.~8, pp. 1450--1455, 2017.

\bibitem{AM1993}
M.~Aoki, M.~Suzuki, H.~Sano, T.~Kawano, T.~Ido, T.~Taniwatari, K.~Uomi, and
  A.~Takai, ``Ingaas/ingaasp mqw electroabsorption modulator integrated with a
  dfb laser fabricated by band-gap energy control selective area mocvd,''
  \emph{IEEE journal of quantum electronics}, vol.~29, no.~6, pp. 2088--2096,
  1993.

\bibitem{KS2015}
S.~Koeber, R.~Palmer, M.~Lauermann, W.~Heni, D.~L. Elder, D.~Korn, M.~Woessner,
  L.~Alloatti, S.~Koenig, P.~C. Schindler \emph{et~al.}, ``Femtojoule
  electro-optic modulation using a silicon--organic hybrid device,''
  \emph{Light: Science \& Applications}, vol.~4, no.~2, pp. e255--e255, 2015.

\bibitem{LM2002}
M.~Lee, H.~E. Katz, C.~Erben, D.~M. Gill, P.~Gopalan, J.~D. Heber, and D.~J.
  McGee, ``Broadband modulation of light by using an electro-optic polymer,''
  \emph{Science}, vol. 298, no. 5597, pp. 1401--1403, 2002.

\bibitem{HC2018}
C.~Haffner, D.~Chelladurai, Y.~Fedoryshyn, A.~Josten, B.~Baeuerle, W.~Heni,
  T.~Watanabe, T.~Cui, B.~Cheng, S.~Saha \emph{et~al.}, ``Low-loss
  plasmon-assisted electro-optic modulator,'' \emph{Nature}, vol. 556, no.
  7702, pp. 483--486, 2018.

\bibitem{HM2019}
M.~He, M.~Xu, Y.~Ren, J.~Jian, Z.~Ruan, Y.~Xu, S.~Gao, S.~Sun, X.~Wen, L.~Zhou
  \emph{et~al.}, ``High-performance hybrid silicon and lithium niobate
  mach--zehnder modulators for 100 gbit s- 1 and beyond,'' \emph{Nature
  Photonics}, vol.~13, no.~5, pp. 359--364, 2019.

\bibitem{WX2019}
X.~Wang, P.~O. Weigel, J.~Zhao, M.~Ruesing, and S.~Mookherjea, ``Achieving
  beyond-100-ghz large-signal modulation bandwidth in hybrid silicon photonics
  mach zehnder modulators using thin film lithium niobate,'' \emph{APL
  Photonics}, vol.~4, no.~9, p. 096101, 2019.

\bibitem{AA2020}
A.~N.~R. Ahmed, S.~Shi, A.~Mercante, S.~Nelan, P.~Yao, and D.~W. Prather,
  ``High-efficiency lithium niobate modulator for k band operation,'' \emph{Apl
  Photonics}, vol.~5, no.~9, p. 091302, 2020.

\bibitem{JM2021}
M.~Jin, J.~Chen, Y.~Sua, P.~Kumar, and Y.~Huang, ``Efficient electro-optical
  modulation on thin-film lithium niobate,'' \emph{Optics Letters}, vol.~46,
  no.~8, pp. 1884--1887, 2021.

\bibitem{SR2020}
R.~Safian, M.~Teng, L.~Zhuang, and S.~Chakravarty, ``Foundry-compatible thin
  film lithium niobate modulator with rf electrodes buried inside the silicon
  oxide layer of the soi wafer,'' \emph{Optics Express}, vol.~28, no.~18, pp.
  25\,843--25\,857, 2020.

\bibitem{CS2021}
S.~Chakravarty, M.~Teng, R.~Safian, and L.~Zhuang, ``Hybrid material
  integration in silicon photonic integrated circuits,'' \emph{Journal of
  Semiconductors}, vol.~42, no.~4, p. 041303, 2021.

\bibitem{LX2020}
X.~Liu, B.~Xiong, C.~Sun, Z.~Hao, L.~Wang, J.~Wang, Y.~Han, H.~Li, J.~Yu, and
  Y.~Luo, ``Low half-wave-voltage thin film linbo3 electro-optic modulator
  based on a compact electrode structure,'' in \emph{Asia Communications and
  Photonics Conference}, 2020, pp. M4A--144.

\bibitem{LH2017}
H.~Liang, R.~Luo, Y.~He, H.~Jiang, and Q.~Lin, ``High-quality lithium niobate
  photonic crystal nanocavities,'' \emph{Optica}, vol.~4, no.~10, pp.
  1251--1258, 2017.

\bibitem{WJ2015}
J.~Wang, F.~Bo, S.~Wan, W.~Li, F.~Gao, J.~Li, G.~Zhang, and J.~Xu, ``High-q
  lithium niobate microdisk resonators on a chip for efficient electro-optic
  modulation,'' \emph{Optics express}, vol.~23, no.~18, pp. 23\,072--23\,078,
  2015.

\bibitem{ZM2021}
M.~Zhang, C.~Wang, P.~Kharel, D.~Zhu, and M.~Lon{\v{c}}ar, ``Integrated lithium
  niobate electro-optic modulators: when performance meets scalability,''
  \emph{Optica}, vol.~8, no.~5, pp. 652--667, 2021.

\bibitem{WP2020}
P.~O. Weigel, F.~Valdez, J.~Zhao, H.~Li, and S.~Mookherjea, ``Design of
  high-bandwidth, low-voltage and low-loss hybrid lithium niobate electro-optic
  modulators,'' \emph{Journal of Physics: Photonics}, vol.~3, no.~1, p. 012001,
  2020.

\bibitem{SS2021}
S.~Sun, M.~Xu, M.~He, S.~Gao, X.~Zhang, L.~Zhou, L.~Liu, S.~Yu, and X.~Cai,
  ``Folded heterogeneous silicon and lithium niobate mach--zehnder modulators
  with low drive voltage,'' \emph{Micromachines}, vol.~12, no.~7, p. 823, 2021.

\bibitem{RA2016}
A.~Rao, A.~Patil, P.~Rabiei, A.~Honardoost, R.~DeSalvo, A.~Paolella, and
  S.~Fathpour, ``High-performance and linear thin-film lithium niobate
  mach--zehnder modulators on silicon up to 50 ghz,'' \emph{Optics letters},
  vol.~41, no.~24, pp. 5700--5703, 2016.

\bibitem{CL2014}
L.~Chen, Q.~Xu, M.~G. Wood, and R.~M. Reano, ``Hybrid silicon and lithium
  niobate electro-optical ring modulator,'' \emph{Optica}, vol.~1, no.~2, pp.
  112--118, 2014.

\bibitem{YY2022}
Y.~Yamaguchi, P.~T. Dat, S.~Takano, M.~Motoya, Y.~Kataoka, J.~Ichikawa,
  S.~Oikawa, R.~Shimizu, N.~Yamamoto, A.~Kanno \emph{et~al.}, ``Low-loss
  ti-diffused linbo 3 modulator integrated with electro-optic frequency-domain
  equalizer for high bandwidth exceeding 110 ghz,'' in \emph{2022 European
  Conference on Optical Communication (ECOC)}, 2022, pp. 1--4.

\bibitem{GG2009}
G.~Ghione, \emph{Semiconductor devices for high-speed optoelectronics}.\hskip
  1em plus 0.5em minus 0.4em\relax Cambridge University Press Cambridge, 2009,
  vol. 116.

\end{thebibliography}

\end{document}